\documentclass[aps,prd,superscriptaddress,floatfix]{revtex4-1}
\usepackage{color}
\usepackage{amssymb}
\usepackage[normalem]{ulem}
\usepackage{graphicx,color}
\usepackage{amsmath}
\usepackage{subfigure}
\usepackage[normalem]{ulem}
\usepackage{booktabs}
\usepackage{xcolor}
\usepackage{epsfig,graphics,color,graphicx,amsmath}

\colorlet{darkgreen}{green!50!black}
\colorlet{brightyellow}{yellow!75!red}
\colorlet{orange}{red!50!yellow}
\colorlet{darkblue}{blue!60!black}
\colorlet{darkred}{red!80!black}

\usepackage{soul}

\usepackage{mathtools}
\usepackage{mathrsfs}
\usepackage{bm}
\usepackage{relsize}
\usepackage{indentfirst}

\usepackage{xcolor}

\usepackage{amssymb,amsmath,multirow,epsfig,graphicx,color}

\usepackage{epsfig}
\usepackage{graphicx,amsmath}
\def\be{\begin{eqnarray} &&}

	\def\ee{\end{eqnarray}}

\newcommand\ba{\begin{eqnarray}}
	\newcommand\ea{\end{eqnarray}}

\newcommand{\bas}{\begin{eqnarray*}}
	\newcommand{\eas}{\end{eqnarray*}}

\newcommand{\bno}{\begin{eqnarray*}}
	\newcommand{\eno}{\end{eqnarray*}}

\def\sl
\usepackage{hyperref}
\usepackage{url}
\usepackage{hyperref}
\hypersetup{
	colorlinks=true,  
	linkcolor=blue,   
	filecolor=magenta,      
	urlcolor=blue,    
	citecolor=blue,   
} 
\begin{document}
	\vspace{-12ex}    
	\begin{flushright} 	
		{\normalsize \bf \hspace{50ex}}
	\end{flushright}
	\vspace{11ex}
	\title{A Quadratic Equation of State to Cosmic Acceleration:\\ Entropy Evolution and Phantom Crossing}
	\author{Ashraf Shahriar}
	\email{ashraf.shahriar@yahoo.com}
	\affiliation{Department of Physics, Qo. C., Islamic Azad University, Qom, Iran}
	\author{Mohammad Abbasiyan-Motlaq}
	\email{ma.motlaq@gmail.com (Corresponding Aouthor)}
	\affiliation{Department of Physics, Qo. C., Islamic Azad University, Qom, Iran}
	\author{Majid Mohsenzadeh}
	\email{ma.mohsenzadeh@iau.ac.ir}
	\affiliation{Department of Physics, Qo. C., Islamic Azad University, Qom, Iran}
	\author{Ebrahim Yusofi}
	\email{eyusofi@ipm.ir}

	\affiliation{Innovation and Management Research Center, AA. C., Islamic Azad University, Amol, Iran}
	\affiliation{School of Astronomy, Institute for Research in Fundamental Sciences(IPM), P. O. Box 19395-5531,Tehran, Iran}
	
	\date{\today}

\begin{abstract}
	This paper investigates the thermodynamic evolution of the universe within the framework of a quadratic equation of state (EoS). Building upon the basis of the quadratic EoS model, as a phenomenological extension to dark energy models, we analyze the implications for cosmic dynamics, including energy density evolution of effective dark matter and dark energy, entropy behavior, and convexity stability conditions. Our approach emphasizes the significance of thermodynamic principles in understanding the late-time acceleration and the crossing of the phantom divide, providing a cohesive description consistent with recent observational data. Moreover, we demonstrate that the $\Lambda$CDM model, regardless of entropy additivity, violates the convexity condition, while the quadratic model aligns with maximum entropy and may prevent a \textit{Big Rip} scenario.
		\\
		\noindent \hspace{0.35cm} 
		\noindent \hspace{0.35cm}  	
	\end{abstract}

	\maketitle
\section{Introduction}
Recent observations, such as type Ia supernovae (SN), cosmic microwave background (CMB), large-scale structure (LSS) of galaxies, and baryon acoustic oscillations (BAO), reveal that the universe's expansion is accelerating, necessitating a mysterious component called dark energy~\cite{Supernova:1998vns, SupernovaTeam:1998fmf, Planck:2018vyg, SDSS:2003eyi, WMAP:2003elm, SDSS:2005xqv}. The cosmological constant ($\Lambda$), with an equation of state parameter $\omega_{\Lambda} = -1$, offers a leading explanation for this acceleration and is central to the $\Lambda$CDM model, which aligns with many cosmological observations~\cite{Peebles:2002gy,1993Peeblesbook}. However, this model encounters significant $H_0$ and $S_8$ tensions, notably between the Hubble constant derived from CMB data ($H_0 \simeq 67$ $\text{km}$ $\text{s}^{-1}$ $\text{Mpc}^{-1}$) and the Cepheid-calibrated value ($H_0 \simeq 73 $ $\text{km}$ $\text{s}^{-1}$ $\text{Mpc}^{-1}$)~\cite{Riess:2020fzl, Riess:2019cxk, DiValentino:2020zio,Vagnozzi:2023nrq}, as well as discrepancies in the amplitude of matter fluctuations inferred from CMB ($S_8 \simeq 0.832$) versus LSS data ($S_8 \simeq 0.762$)~\cite{Joudaki:2019pmv, KiDS:2020suj, DES:2021bvc}. Recent analyses from the Atacama Cosmology Telescope (ACT) indicate that the $S_8$ tension appears to be confined to the late universe ($z \leq 2$), suggesting that the observed discrepancies are more significant at lower redshifts and may be related to late-universe physics rather than early-universe conditions~\cite{Akarsu:2024qiq}. Also, the Standard Model fails to adequately explain dark matter, dark energy, and their origins. These issues motivate physicists to investigate alternatives to the standard $\Lambda$CDM model and new physics for early and late time cosmology~\cite{Vagnozzi:2023nrq}.
Nevertheless, the most recent and formidable challenge to the \(\Lambda\)CDM model of cosmology is the remarkable consistency of DESI's observations with models featuring evolving dark energy, particularly given the indications of a crossing of the phantom divide line (\(w = -1\)), which conflicts with the simplest cosmological constant scenarios ~\cite{DESI:2025zgx, DESI2023, Wang2022}.

To address the physical challenges in cosmology, numerous theoretical models have been developed~\cite{CosmoVerse:2025txj}. Among these, some focus on modifying the curvature term in the Einstein field equations, known as modified gravity models \cite{Odintsov:2020qzd, Schiavone:2022wvq,Adil:2021zxp, DiValentino:2021izs, Moshafi:2020rkq}. Others adjust the energy-momentum tensor, referred to as modified dark energy models ~\cite{Pan:2019gop, Karwal:2021vpk, Niedermann:2020dwg, Li:2019yem}. Many of these modified dark energy models expand upon the $\Lambda$CDM framework. For instance, dynamical dark energy models permit interactions between dark energy and dark matter \cite{Dahmani:2023bsb}, and early dark energy models behave similarly to a cosmological constant in the early universe \cite{Li:2019yem}. Some of these approaches may help alleviate the $H_0$ and $S_8$ tension~\cite{DiValentino:2021izs}.

Another promising and impactful avenue of research lies in the generalization of the linear equation of state (EoS) to nonlinear forms. In particular, quadratic EoS models have recently garnered renewed and increasing attention, highlighting their potential to unlock deeper insights into cosmic dynamics. Remarkably, despite their contemporary resurgence, the origins of these models can be traced back to foundational studies, underscoring their enduring relevance and the profound importance of exploring nonlinear EoS approaches in advancing our understanding of the universe. Barrow \cite{Barrow:1990vx} was among the first to derive analytical solutions to the Friedmann equations for EoS of the form $f(\rho) = -\rho + \alpha \rho^{q}$, illustrating their relationship to scalar fields featuring specific self-interactions capable of driving inflation. Subsequent investigations have examined various facets of quadratic EoS:

\begin{itemize}
	\item Nojiri and Odintsov \cite{Nojiri:2004pf} analyzed scale factors with evolving deceleration parameters and derived the corresponding quadratic EoS.
	\item Stefanci\'c and Alcaniz \cite{Stefancic:2004kb} explored potential future singularities, such as the big rip, within Barrow's framework.
	\item Ananda and Bruni \cite{Ananda:2006gf} systematically classified the dynamical behaviors of universes modeled by Robertson-Walker and Bianchi I geometries with quadratic EoS.
	\item Chavanis \cite{Chavanis:2012gp} presented comprehensive analyses of quadratic EoS, including their connections to Bose-Einstein condensates as dark matter candidates.
	\item Berteaud et al. \cite{Berteaud:2018ifl} focused on a particular case of quadratic EoS, providing significant physical insight with minimal complexity. This model presents a compelling alternative to traditional dark energy scenarios.
\end{itemize}
In the present work, we adopt a quadratic equation of state as,
\begin{equation}
	\label{cvt3}
	P = w \rho + b \rho^2,
\end{equation}
to model the combined influence of effective dark energy and dark matter in cosmic dynamics. Within an effective cosmic web, we derive a generalized Hubble parameter and quadratic equation of state for two coexisting components: a high-density phase dominated by effective dark matter and a low-density phase dominated by effective dark energy. Additionally, we evaluate the validity of the generalized second law of thermodynamics and the entropy maximization condition within this framework, comparing the results with those obtained in the standard \(\Lambda\)CDM model and the latest DESI DR2 data~\cite{DESI:2025zgx}.

The paper is organized as follows. In Section \ref{II}, we introduce a modified cosmic fluid that incorporates a quadratic term complementing the linear term in the equation of state. Subsequent subsections \ref{B.} through \ref{D.} analyze the dynamics of this model using the effective equation of state parameter and assess its consistency with observational constraints. Section \ref{III} focuses on the Hubble parameter and its evolution within the proposed framework. In Section \ref{IV}, we evaluate the model’s validity through entropy considerations, examining the entropy and its first and second derivatives in accordance with the Generalized Second Law (GSL). Section \ref{V} investigates the maximum entropy condition for the standard 
$\Lambda$CDM model in both additive and non-additive scenarios, demonstrating that the convexity condition does not hold in these cases. Using the effective dark energy component in our model, we address this limitation and compare our results with recent observational constraints on the dark energy equation of state, including the phantom crossing scenario. Finally, Section \ref{VI} summarizes the proposed model and discusses its potential implications.

\section{Quadratic Cosmic Fluid}
\label{II}
Traditional cosmological models often treat the cosmic fluid as a perfectly homogeneous and isotropic entity characterized by a simple linear equation of state (EoS). However, this approximation neglects the nonlinear effects that dynamically alter the fluid's effective thermodynamic behavior.

In this work, we propose an effective EoS that explicitly incorporates these nonlinearities through a second-order (quadratic) term, enhancing the model’s capacity to represent the influence of high-density structures on large-scale dynamics. In our effective quadratic EoS \eqref{cvt3}, we interpret the term \(w\rho\) as capturing the linear, average behavior of the cosmic fluid, aligning with the standard matter or dark energy components. Meanwhile, the quadratic term \(b\rho^2\) represents correctional effects that become significant at high densities, reflecting the influence of dense structures in the universe.
Quadratic EoS (e.g., $P = w\rho + b\rho^2$) have been explored as phenomenological extensions to dark energy models ~\cite{Barrow:1990vx,Nojiri:2004pf,Stefancic:2004kb,Ananda:2006gf,Chavanis:2012gp,Chavanis:2012gp,Berteaud:2018ifl,Mohammadi:2023yms, Moshafi:2024guo, Buchert:2022zaa,Ananda:2006gf, Capozziello:2005pa, Bento:2002ps, Mohammadi:2023yms}, offering flexibility to address observational tensions like phantom crossing.
\subsection{Effective Energy Density for Quadratic Model}
\label{B.}
 For the quadratic equation of state (\ref{cvt3}) characterized by parameters $w$ and $b$, the energy density evolves as~\cite{Mohammadi:2023yms},

\begin{equation}
	\label{rho_evolution}
	\rho(\xi,w,z) = \frac{\rho_0 (1+z)^{3(1+w)}}{1 + \frac{\xi}{1+w} \left[ 1 - (1+z)^{3(1+w)} \right]},
\end{equation}

where $\rho_0$ denotes the current (at $z=0$) density, and the dimensionless parameter $\xi$ encapsulates the effects of nonlinear structures, defined as $\xi = b \rho_0$, with $b$ representing a characteristic scale of the structure formation processes. 

\subsection{Effective Equation of State Parameter for Quadratic Model}
\label{C.}

The equation of state parameter $w$ in cosmology quantifies the pressure-to-energy density ratio of a cosmic component, defining its thermodynamic behavior (e.g., matter, radiation, or dark energy). The generalized form of the equation of state parameter can be written as
\begin{equation}
	\label{eq:eos_def}
	w_{\rm x} = \frac{P_{\rm x}}{\rho_{\rm x}},
\end{equation}

For the quadratic case (\ref{cvt3}), we obtain from \eqref{eq:eos_def},
\begin{equation}
	\label{eq:eos_quadratic}
	w_{\rm x}(\rho) = w_{\rm x} + b\rho_{\rm x}(\xi_{\rm x},w_{\rm x}, z),
\end{equation}

Assuming the current universe consists primarily of dark matter (dm) and dark energy (de), the effective equation of state for dark energy in the quadratic model can be expressed using Eqs.~\eqref{rho_evolution} and \eqref{eq:eos_quadratic} as
\begin{equation}
	\label{eq:w_de}
	w_{\rm de}(z) = w_{\rm de} + \frac{\xi_{\rm de}(1+z)^{3(1+w_{\rm de})}}{1 + \dfrac{\xi_{\rm de}}{1+w_{\rm de}}\left[1 - (1+z)^{3(1+w_{\rm de})}\right]}.
\end{equation}

Similarly, the equation of state for dynamical dark matter is given by
\begin{equation}
	\label{eq:w_dm}
	w_{\rm dm}(z) = w_{\rm dm} + \frac{\xi_{\rm dm}(1+z)^{3(1+w_{\rm dm})}}{1 + \dfrac{\xi_{\rm dm}}{1+w_{\rm dm}}\left[1 - (1+z)^{3(1+w_{\rm dm})}\right]}.
\end{equation}

Considering the distinct cosmological behaviors of dark matter and dark energy, we assume the pressure generated by dark energy opposes the contraction pressure of dark matter.

Furthermore, the square of the sound speed, defined as $c_s^2 = \partial P / \partial \rho$, must be non-negative to ensure stability,
\begin{equation}
	c_s^2 = w + 2b\rho \geq 0. \label{eq:sound_speed}
\end{equation}

This condition leads to the following implications regarding the sign of the parameters:

\begin{itemize}
	\item If $b > 0$, then to satisfy inequality \eqref{eq:sound_speed}, we require
	\[
	w < 0,
	\]
	which is appropriate for an effective dark energy component characterized by a negative EoS parameter $w_{\rm de} < 0$ and a positive nonlinear coupling coefficient $\xi_{\rm de} > 0$.
	
	\item Conversely, if $b < 0$, then
	\[
	w > 0,
	\]
	aligning with the behavior of an effective dark matter component with $w_{\rm dm} > 0$ and a negative coupling coefficient $\xi_{\rm dm} < 0$.
\end{itemize}
Therefore, we can write,
\begin{equation}
	\label{eq:sign_conventions}
	\underbrace{w_{\rm de}(z) < 0~{\rm and}~ \xi_{\rm de} > 0}\quad, \quad \underbrace{ w_{\rm dm}(z) > 0~{\rm and}~ \xi_{\rm dm} < 0}.
\end{equation}

\subsection{Special Cases and Consistency with Observations}
\label{D.}
For the special case of a dark energy dominated fluid with \( w_{\rm de} = w_{\Lambda} = -1 \), the equation of state derived from (\ref{eq:w_de}) simplifies to the form~\cite{Mohammadi:2023yms},
\begin{equation}
	\label{wdezsp}
	w_{\rm de}(z) = -1 + \frac{\xi_{\rm de}}{1 - 3 \xi_{\rm de} \ln(1+z)},
\end{equation}
which describes a dynamical dark energy component rather than a strict cosmological constant with fixed density \( \rho_{\Lambda} \). Unlike the constant vacuum energy density, which remains invariant during the universe's evolution, this model allows the effective dark energy equation of state \( w_{\rm de}(z) \) to vary with the expansion history, providing a flexible framework to accommodate observational data \cite{Tripathi:2016slv}.

Furthermore, this functional form naturally facilitates a crossing of the so-called phantom divide (\( w = -1 \)), enabling \( w_{\rm de}(z) \) to evolve into super-negative, zero, or even positive regimes depending on the sign and magnitude of the free parameter \( \xi_{\rm de} \). In particular, appropriate choices of \( \xi_{\rm de} \) can reconcile the model with recent results from the DESI DR2 survey \cite{DESI:2025zgx}, such as the possibility of the effective dark energy equation of state shifting toward values greater than \(-1\), approaching zero, or becoming positive, as illustrated in Fig. (12) of \cite{DESI:2025zgx}.

Also, for the dark matter component of the universe with $w_{\rm dm}=0$, the energy density depends on $z$ similar to interacting non-cold dark matter~\cite{Pan:2022qrr} as follows,
\begin{equation}
	\label{wdmzsp}
	w_{\rm dm}(z)=  
	\frac{\xi_{\rm dm}(1+z)^{3}}{1+\xi_{\rm dm}[1-(1+z)^{3}]}	\end{equation}
In cosmology, dark matter is commonly assigned an equation of state parameter $w$ that is close to zero, as dark matter is non-relativistic and lacks strong interactions that would impact its pressure significantly. This assumption reflects dark matter's behavior as non-interacting cold particles, exerting gravitational effects on the universe without making a substantial contribution to its overall pressure. However, in the new relations (\ref{wdezsp}) and (\ref{wdmzsp}) for the dark matter (energy) equation of state, we derived additional terms involving the nonlinear process ($\xi \neq 0$). In cosmology, these terms introduce an additional (deficit) pressure that can be considered as a potential new source for effective dark matter and dark energy.

Only in the non-quadratic case where $\xi_{\rm de} = 0$ and $\xi_{\rm dm} = 0$, the equations (\ref{wdezsp}) and (\ref{wdmzsp}) align with the equations of state for the cosmological constant and the non-interacting dark matter, respectively. Therefore, (\ref{wdezsp}) and (\ref{wdmzsp}) can represent a generalized form of the standard model with consideration of nonlinear process.

We can obtain, at the limits $z\rightarrow0$ for effective dark energy as below, 
\begin{equation}
	\label{slozz}
	w_{\rm de}(z\rightarrow0)=-1+ \xi_{\rm de}~, 
\end{equation}
and for effective dark matter we obtain,
\begin{equation}
	\label{sloz}
	w_{\rm dm}(z\rightarrow0)= \xi_{\rm dm}~,
\end{equation}
Given the flexibility of the factor $\xi_x$ to accept both positive and negative values, the equations of state for dark matter and dark energy can not only vary in magnitude but also in sign. Notably, relation (\ref{sloz}) indicates that the equation of state for dark matter could be non-zero or even negative. This scenario could potentially address the Hubble tension, $\sigma_8$ discrepancies, and the ISW-void anomaly simultaneously \cite{Akarsu:2024qiq, Naidoo:2023tpz}. The redshift dependence of 
$\sigma_8 (S_8) $, interpolating through the Planck value, mirrors observations in the ISW anomaly \cite{Akarsu:2024qiq}. 

The latest data from CMB observations, particularly from the Planck~\cite{Planck:2018jri}, indicate that \( w_{\rm de} \) can be more accurately estimated as,
\be
\label{wobs}
w_{\rm de}^{\mathrm{planck}} = -1.03 \pm 0.03 
\ee
Also, a combination of Planck CMB data, baryon acoustic oscillations, type Ia supernovae, and cosmic chronometer data all agree with a final value of EoS parameter as~\cite{Escamilla:2023oce, Yusofi:2018lqb} 
\be
\label{wobs2}
w_{\rm de}^{\rm combine} = -1.013^{+0.038}_{-0.043}~.
\ee
Also according to an interacting non-cold dark matter scenario suggested that investigating the possibility of non-cold dark matter in the universe is worth exploring
further to gain a better understanding of the nature of dark matter. In \cite{Kumar:2012gr, Pan:2022qrr}, the best fit values with a combination of CMB$+$Lensing$+$BAO$+$Pantheon is reported as,
\be
\label{wobs2}
w_{\rm dm}^{\rm combine} = 0.00108^{+0.00040}_{-0.00096}< 0.0023~.
\ee
More precisely in present universe in redshift $z\rightarrow0$, based on the observational data, we may adopt the approximations \( w_{\rm dm} = 0 + \xi_{\rm dm} \) and \( w_{\rm de} = -1 + \xi_{\rm de} \), where $\xi_{\rm de}\sim 10 \times \xi_{\rm dm}$ (\( 0 < \xi_{\rm de} < 0.1 \)).

\section{Hubble Expansion for Quadratic Model}
\label{III}
\begin{table}[t]
	\centering
	\caption{Summary of Model Variants}
	\label{tab:models}
	\renewcommand{\arraystretch}{1.5}
	\begin{tabular}{|c|c|}
		\hline
		\textbf{Model} & \textbf{Description} \\
		\hline
		\textbf{EMEM} & Both Effective Dark Matter and Dark Energy (with quadratic pressure correction $\xi_{\rm dm}\neq 0$ and $\xi_{\rm de}\neq 0$) \\
		\hline
		\textbf{EMDM} & Effective Dark Matter Dominated Model (with quadratic pressure correction $\xi_{\rm dm}\neq 0$)\\
		\hline
		\textbf{EEDM} & Effective Dark Energy Dominated Model (with quadratic pressure correction $\xi_{\rm de}\neq0$) \\
		\hline
		\textbf{$w$CDM} & Non-Effective Case (without quadratic pressure correction $\xi_{\rm dm} = \xi_{\rm de} = 0$, $w_{\rm dm} = 0$) \\
		\hline
		\textbf{$\Lambda$CDM} & Standard Model (without quadratic pressure correction $\xi_{\rm dm} = \xi_{\rm de} = 0$, $w_{\rm dm} = 0$, $w_{\rm de} = -1$) \\
		\hline
	\end{tabular}
\end{table}
\begin{itemize}
	
	\item {\textbf{Model I (EMEM)}}:
The generalized form of the Hubble parameter in our quadratic model, which takes into account the effective energy density for DM and DE, is defined as (assuming \( \xi_{\rm dm}\), and  $~\xi_{\rm de}$ are positive values),

\begin{align}
	\label{HMCV}
	H_{\rm EMEM} = H(\xi_{\rm dm}, \xi_{\rm de}, w_{dm}, w_{de}, z) = H_{\rm 0} \left[ \frac{\Omega^{0}_{dm} (1+z)^{3(1+w_{dm})}}{1 \pmb{-} \frac{\xi_{\rm dm}}{1+w_{dm}} \left(1 - (1+z)^{3(1+w_{dm})}\right)} + \frac{\Omega^{0}_{de} (1+z)^{3(1+w_{de})}}{1 \pmb{\textbf{+}} \frac{\xi_{\rm de}}{1+w_{de}} \left(1 - (1+z)^{3(1+w_{de})}\right)} \right]^{\frac{1}{2}}.
\end{align}
\begin{figure}[ht]
	\centering
	\includegraphics[width=16 cm]{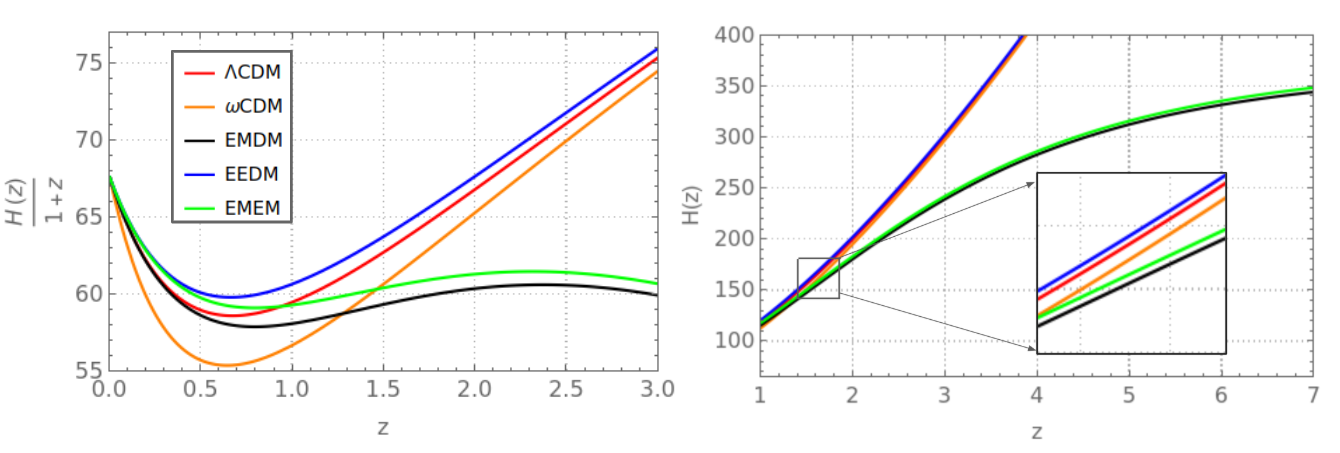}
	\caption{\footnotesize The evolution rate of the scale factor $\dot{a}=\frac{H(z)}{1+z}$ (left) and Hubble parameter $H(z)$ (right) are shown for five different models as a function of redshift $z$. The plot utilizes parameters $w_{\rm de} = -1.03$, $w_{\rm dm} = 0.003$, $\xi_{\rm dm} = 0.1$, and $\xi_{\rm dm} = 0.01$.} 
	\label{HUB}
\end{figure}
\begin{figure}[ht]
	\centering
	\includegraphics[width=16 cm]{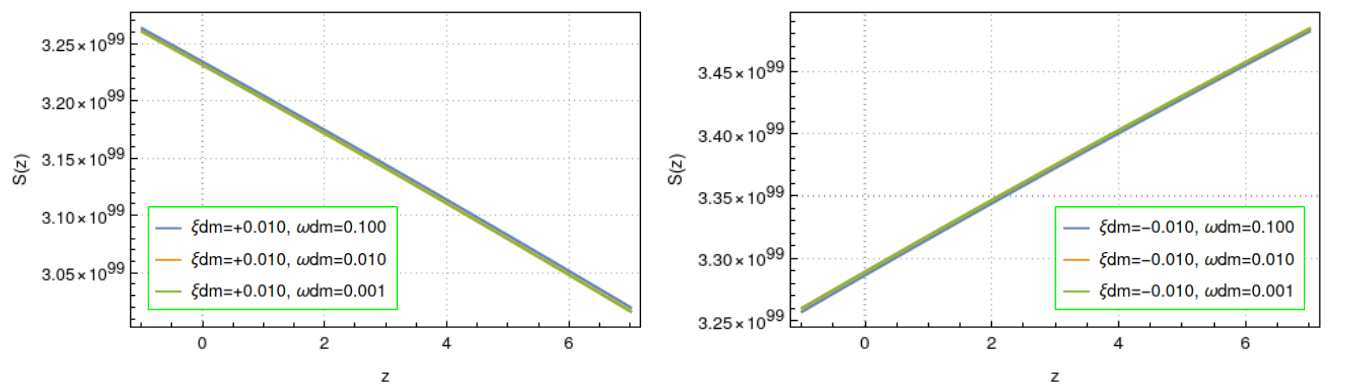}
	\includegraphics[width=16 cm]{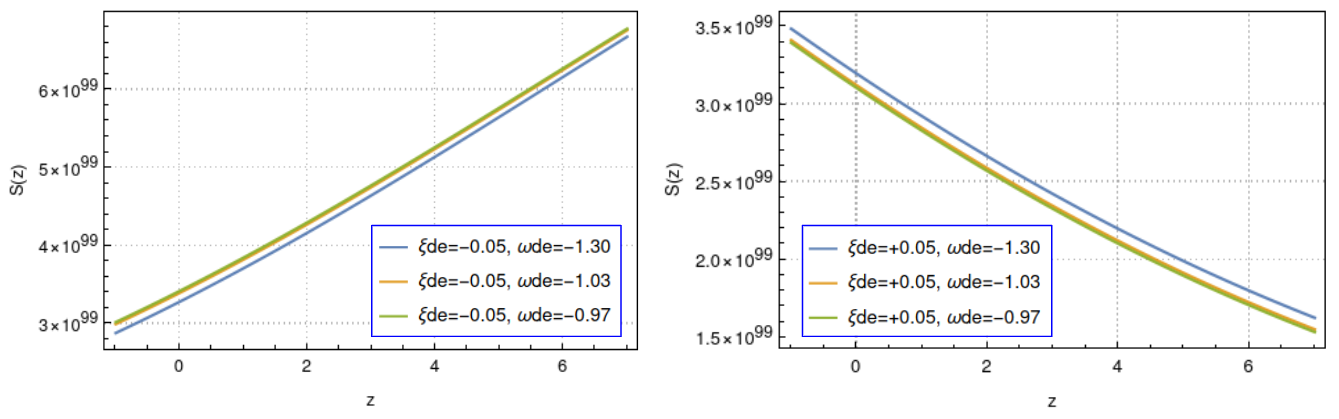}
	\caption{\footnotesize  \textbf{Sign selection for the effective dark matter and dark energy through entropy:} Utilizing opposite signs for $\xi_{\rm dm}$ and $w_{\rm dm}$ results in accurate negative entropy changes, as depicted in the right plot. In contrast, using the same signs leads to an increase in entropy, shown in the left plot (Top row for model II). When \( \xi_{\rm de} \) and \( w_{\rm de} \) share the same sign, it generates false negative entropy changes (left plot). Conversely, employing opposite signs yields an increase in entropy, as illustrated in the right plot (Bottom row for model III). }
	\label{SIGN}
\end{figure}
The minus sign ($\pmb{-}$) before the $\xi_{\rm dm}$ and the positive one ($\pmb{+}$) before the $\xi_{\rm de}$ associated  with conditions (\ref{eq:sign_conventions}). As shown in Fig.\ref{HUB}, models that incorporate dominated effective dark matter result in a lower Hubble rate, while those that include effective dark energy yield higher values of the Hubble parameter. 
	\item {\textbf{Model II (EMDM)}}:
For the universe that only dominated with effective dark matter (\( \xi_{\rm dm} \neq 0 \) and \( \xi_{\rm de} = 0 \)), we derive from equation (\ref{HMCV}):

\begin{align}
	\label{HMC}	
	H_{\rm EMDM} = H(\xi_{\rm dm},0 , w_{dm}, w_{de}, z) = H_{\rm 0} \left[ \frac{\Omega^{0}_{dm} (1+z)^{3(1+w_{dm})}}{1 - \frac{\xi_{\rm dm}}{1+w_{dm}} \left(1 - (1+z)^{3(1+w_{dm})}\right)} + \Omega^{0}_{de} (1+z)^{3(1+w_{de})} \right]^{\frac{1}{2}}.
\end{align}
	\item {\textbf{Model III (EEDM)}}:
In the case of our model of the universe that dominated only with effective dark energy (\( \xi_{\rm de} \neq 0 \) and \( \xi_{\rm dm} = 0 \)), we can reformulate equation (\ref{HMCV}) as:

\begin{align}
	\label{HMVD}
	H_{\rm EEDM} = H(0, \xi_{\rm de}, w_{dm}, w_{de}, z) = H_{\rm 0} \left[ \Omega^{0}_{de} (1+z)^{3(1+w_{dm})} + \frac{\Omega^{0}_{de} (1+z)^{3(1+w_{de})}}{1 + \frac{\xi_{\rm de}}{1+w_{de}} \left(1 - (1+z)^{3(1+w_{de})}\right)} \right]^{\frac{1}{2}}.
\end{align}


	\item {\textbf{Model IV ($w$CDM)}}: Furthermore, in the non-quadratic case ($ \xi_{\rm dm} = 0, ~ \xi_{\rm de} = 0 $), with ($w_{\rm dm} = 0$ and $ w_{\rm de}  \neq w_{\Lambda} = -1$), we receive $w$CDM case,
\begin{align}
	\label{HMV}
	H_{w\rm CDM} = H(0, 0, 0, w_{de}, z) = H_{\rm 0} \left[ \Omega^{0}_{dm} (1+z)^{3} +\Omega^{0}_{de} (1+z)^{3(1+w_{de})} \right]^{\frac{1}{2}}.
\end{align}
	\item {\textbf{Model V (\(\Lambda\)CDM)}}: Finally with ($w_{dm} = 0 \) and \( w_{de} = w_{\Lambda} = -1$), we can recover the standard \(\Lambda\)CDM model from equation (\ref{HMCV}), yielding the familiar Hubble parameter:

\[
H_{\rm \Lambda CDM} = H(0, 0, 0,-1, z) = H_{\rm 0} \left[ \Omega^{0}_{dm} (1+z)^{3} + \Omega^{0}_{\Lambda} \right]^{\frac{1}{2}}.
\]
\end{itemize}
In Fig.\ref{HUB}, according to the observational data for the equations of state of dark matter and dark energy, we have drawn the Hubble parameter $H(z)$ and the evolution rate of scale factor $\dot{a}$, for different models in comparison with the standard $\Lambda$CDM model. All plots show similar behavior at low redshifts ($z < 1$). However, at higher redshifts, models I (green curve) and II (black curve), which include effective dark matter, diverge from the standard model, displaying significantly different behavior compared to the other models. In contrast, the EEDM model III (blue curve), which incorporates only effective dark energy, aligns more closely with the standard model and yields higher values of the Hubble rate throughout its evolution. To confirm this result, recently, in~\cite{Moshafi:2024guo} the dynamics of EEDM model (with $w_{dm}=0$) were analyzed in background level and linear order of perturbations using data from CMB, BAO, SN, combining these three data (CBS), and a prior on the Hubble constant $H_0$ (R21) to constrain cosmological parameters. The analysis revealed a strong compatibility with R21 and SN, addressing both the \(H_0\) and \(S_8\) tensions simultaneously. However, these tensions remain when combining the datasets and other related plots and tables in Ref. \cite{Moshafi:2024guo}).

\section{Entropy Test of Models}
	\label{IV}
 According to the Generalized Second Law of Thermodynamics (GSL), the total entropy of the universe, which includes both the horizon entropy and the entropy of its interior, is expected to increase over time \cite{Bekenstein:1973ur, Bekenstein:1974ax, Brustein:1999ua}:
	
	\begin{align}
		\label{eqn:GSL}
		\dot{S} = \dot{S}_{\rm H} + \dot{S}_{\rm m} \geq 0,
	\end{align}

In this context, \( S_{\rm m} \) represents the total entropy contributions from all entities that exist within the Hubble horizon, which is the maximum distance from which light can travel to an observer since the beginning of the universe. The dot notation, denoted as \( \dot{S}_{\rm m} \), indicates the rate of change of entropy with respect to cosmic time, reflecting how the entropy evolves as the universe expands.

The entropy contributions included in \( S_{\rm m} \) come from various cosmic components~\cite{KM:2023tyj, Egan:2009yy}:
\begin{itemize}
\item{\textsf{Baryonic Matter (\( S_{\rm b} \)):}}  This refers to the ordinary matter made up of protons, neutrons, and electrons, which is estimated to contribute approximately \( S_{\rm b} \sim 10^{81}k_{\rm B} \), where \( k_{\rm B} \) is the Boltzmann constant.

\item{\textsf{Dark Matter (\( S_{\rm dm} \)):}} This mysterious form of matter, which does not emit or interact with electromagnetic radiation, has an estimated entropy contribution of around \( S_{\rm dm} \sim 10^{88 \pm 1}k_B \).

\item{\textsf{Photons (\( S_{\rm rad} \)):}} The entropy associated with photons, which are particles of light, is estimated to be \( S_{\rm rad} \sim 10^{89}k_{\rm B} \). Photons play a crucial role in the thermal history of the universe, especially during its early stages.

\item{\textsf{Relic Neutrinos (\( S_{\rm \nu} \)):}} These are neutrinos that were produced in the early universe and have persisted to the present day. Their contribution to entropy is estimated at \( S_{\rm \nu} \sim 10^{89}k_{\rm B} \).

\item{\textsf{Relic Gravitons (\( S_{\rm grav} \)):}} Gravitons are hypothetical particles that mediate the force of gravity. The estimated entropy contribution from relic gravitons is around \( S_{\rm grav} \sim 10^{87}k_{\rm B} \).

\item{\textsf{Stellar Black Holes (\( S_{\rm SBH} \)):}} The entropy associated with black holes formed from the collapse of massive stars is estimated to be \( S_{\rm SBH} \sim 10^{97}k_{\rm B} \).

\item{\textsf{Supermassive Black Holes (\( S_{\rm SMBH} \)):}} These are black holes with masses millions to billions of times that of the Sun, typically found at the centers of galaxies. Their entropy contribution is significantly higher, estimated at \( S_{\rm SMBH} \sim 10^{104}k_{\rm B} \).

\item{\textsf{Hubble Horizon (\( S_{\rm H} \)):}} Despite the substantial contributions from these various entities, it is noteworthy that the total entropy associated with the Hubble horizon, denoted as \( S_{\rm H} \), is estimated to be around \( S_{\rm H} \sim 10^{122}k_{\rm B} \). This value vastly surpasses the combined entropy contributions from all of the aforementioned cosmic components. 
\end{itemize}
As a result, in our subsequent discussions, we will primarily focus on the implications and significance of horizon entropy, given its dominance in the overall entropy budget of the universe\cite{Egan:2009yy}.

Additionally, it is important to note that the negative value for \( S''(z) \), which corresponds to the maximum entropy, is anticipated in the future for three different models of cosmic evolution. This suggests that as the universe continues to expand and evolve, the behavior of entropy will exhibit interesting dynamics that could lead to significant implications for our understanding of cosmology and the fate of the universe.

	Bekenstein and Hawking established that the entropy of black holes is intrinsically linked to the area of their event horizon \cite{Bekenstein:1973ur,Bekenstein:1974ax, Hawking:1976bt}. Building on this foundation, Gibbons and Hawking revealed that the entropy associated with the cosmological horizon is also directly proportional to its area. In cosmology, the Hubble horizon serves as a critical framework for assessing horizon entropy, as it defines the observable region at the present epoch. The connection between area and entropy for the observable universe can be expressed mathematically as

\begin{align}
	\label{eqn:SH}
S=S_{\rm H} = \frac{A_{\rm H}}{4l_p^2}k_B,
\end{align}

where \( A_{\text{H}} = 4\pi c^2/H^2 \) is the area of the Hubble horizon, \( l_p \) represents the Planck length, and \( k_B \) denotes the Boltzmann constant. The horizon entropy of effective model can be writed by substituting the expression ($\ref{HMCV}e$) for horizon area into equation (\ref{eqn:SH}), leading to the following in generalized form for the entropy as a function of redshift \( z \),
		\begin{align}
		\label{SMCVM}
		S_{\rm EMEM}(z) = \frac{\pi c^2}{l_p^2 H_{\rm 0}^2}{\left[ \frac{\Omega^{0}_{dm} (1+z)^{3(1+w_{dm})}}{1 \pmb{-} \frac{\xi_{\rm dm}}{1+w_{dm}} \left(1 - (1+z)^{3(1+w_{dm})}\right)} + \frac{\Omega^{0}_{de} (1+z)^{3(1+w_{de})}}{1 \pmb{+} \frac{\xi_{\rm de}}{1+w_{de}} \left(1 - (1+z)^{3(1+w_{de})}\right)}\right]^{-1}}k_{\rm B}~.
		\end{align}
The sign selection for effective terms added to the denominator of fractions in relation (\ref{SMCVM}) follows Fig. \ref{SIGN} and its accompanying caption analysis. The first and second derivative of this entropy with respect to redshift is given by,

\begin{figure}[ht]
	\centering
	\includegraphics[width=\linewidth]{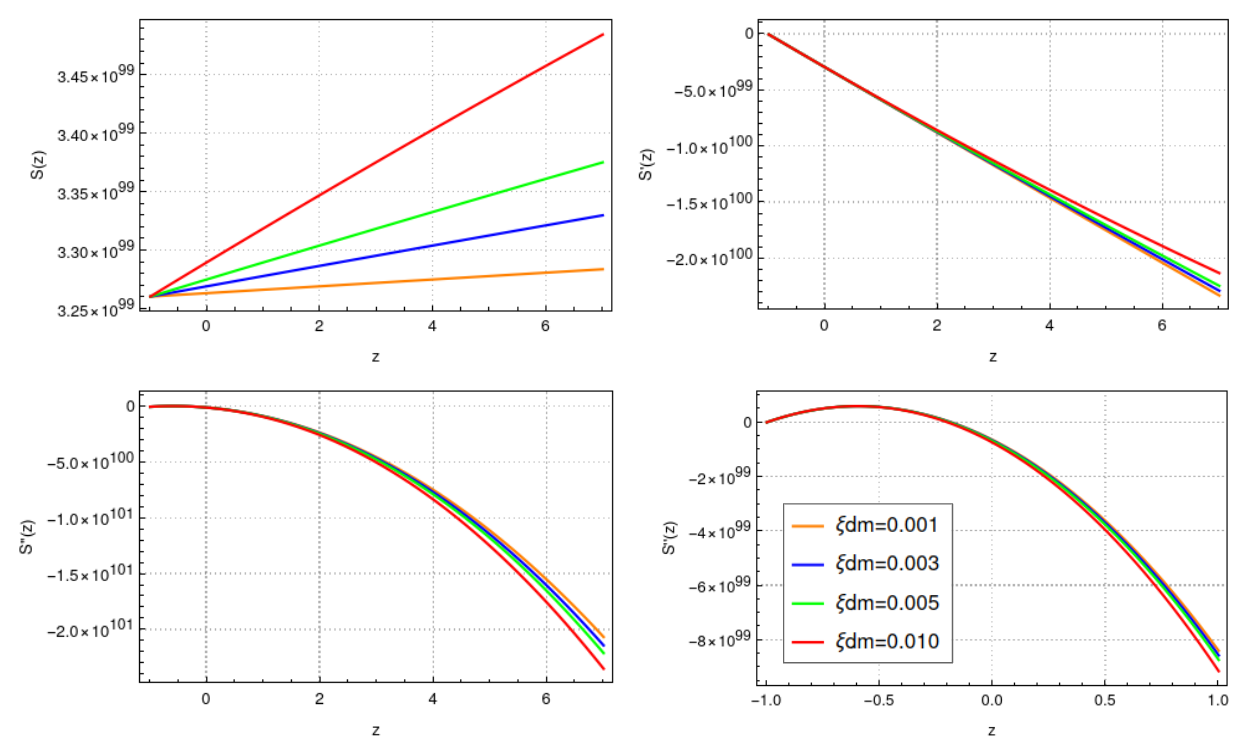}
	\caption{\footnotesize The evolution of entropy \( S(z) \) and its first and second derivatives \( S'(z) \) and \( S''(z) \) for the effective dark matter dominated model (EMDM) is analyzed with respect to redshift. The plots assume \( w_{\rm de} = -1.0003,~  w_{\rm dm} = 0.003, ~ \xi_{\rm de} = 0.0\), and \( \xi_{\rm dm} \leq 0.01 \). Additionally, the non-negative value of \( S''(z) \) for EMDM in the future is shown in the bottom right plot.}
		\label{DDSC}
\end{figure}
\begin{figure}[ht]
	\centering
	\includegraphics[width=\linewidth]{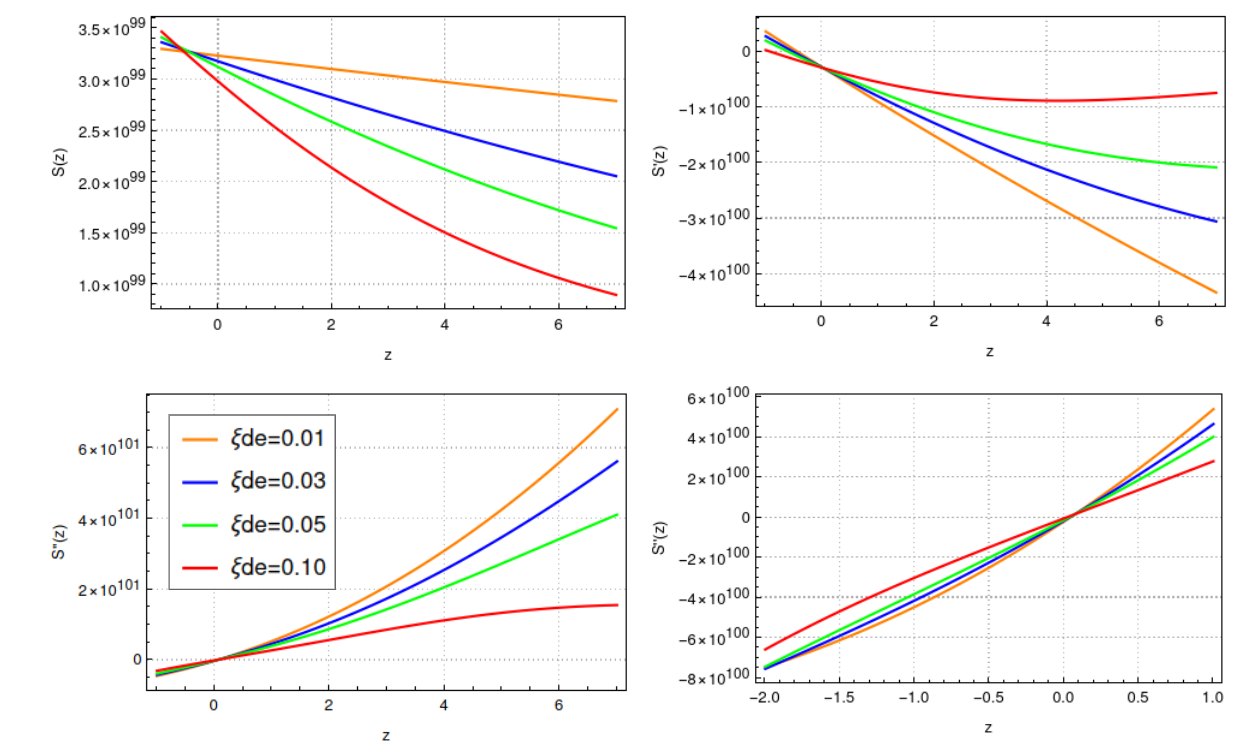}
	\caption{\footnotesize The evolution of entropy \( S(z) \) and its first and second derivatives \( S'(z) \) and \( S''(z) \) for the effective dark energy dominated model (EEDM) is analyzed with respect to redshift. The plots assume \( w_{\rm de} = -1.03,~ w_{\rm dm} = 0.0, ~\xi_{\rm de} \leq 0.1\),~and \(\xi_{\rm dm} = 0.0 \). Additionally, the negative value of \( S''(z) \) for EEDM in the future is shown in the bottom right plot.}
		\label{DDSV}
\end{figure}

\begin{figure}[ht]
	\centering
	\includegraphics[width=\linewidth]{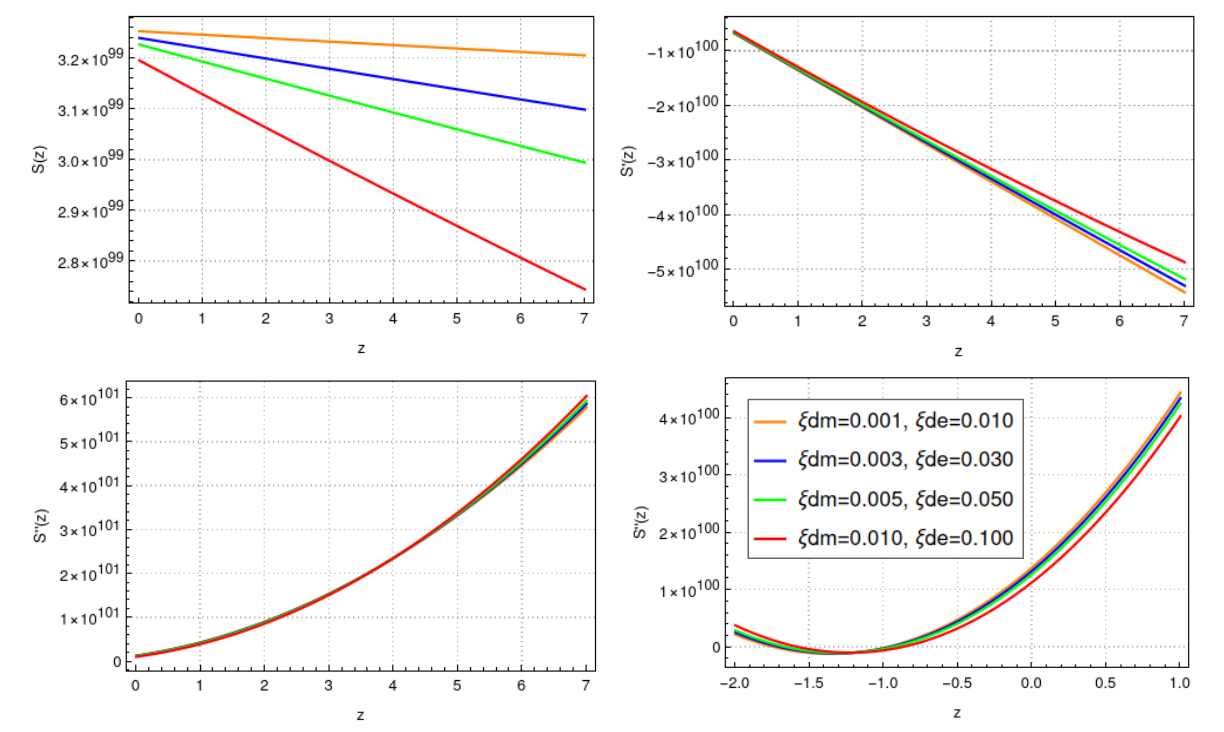}
	\caption{\footnotesize The evolution of entropy \( S(z) \) and its first and second derivatives \( S'(z) \) and \( S''(z) \) for the model includes of both effective dark matter and dark energy (EMEM) is analyzed with respect to redshift. The plots assume \( w_{\rm de} = -1.03,~ w_{\rm dm} = 0.003,~ \xi_{\rm dm} \leq 0.1\),~ and \( \xi_{\rm dm} \leq 0.01 \). Additionally, the negative value of \( S''(z) \) for EMEM in the future is shown in the bottom right plot.}
	\label{DDSCV}
\end{figure}

	\begin{align}
		\label{eqn:SH0}
		S'(z) = \frac{dS}{dz}~\quad,~\qquad S''(z) = \frac{d^{2}S}{dz^2}~.
	\end{align}
Figures \ref{DDSC}, \ref{DDSV}, and \ref{DDSCV} display the entropy and its first and second derivatives for models I, II, and III, respectively. The results indicate that entropy decreases in models involving effective dark matter, while it increases in those with dark energy one, particularly in models where effective dak energy is predominant, the entropy ultimately reaching a maximum.

\section{Maximum entropy and Phantom Crossing}
	\label{V}
The \textit{maximum entropy condition} in cosmological models, such as \(\Lambda\)CDM, suggests that without specific constraints, the universe evolves toward states that maximize entropy, reflecting core thermodynamic and statistical principles. Now this important question arises: \emph{Is achieving maximum entropy in cosmological models a necessity or a choice?}
\begin{itemize}
	\item{\textsf{Necessity:}}
	From a theoretical standpoint, the idea that the universe evolves towards states of maximum entropy aligns with the second law of Thermodynamics, which states that in an isolated system, entropy tends to increase over time. Cosmologically, as the universe expands and evolves, one might argue that reaching a state of maximum entropy could be seen as a necessary outcome. This would imply that certain characteristics of the universe and its ultimate fate (Thermodynamic equilibrium and avoid of Big Rip in our model) are dictated by this tendency towards maximum entropy.
	\item{\textsf{Choice:}}
	On the other hand, the application of maximum entropy principles in model selection could be viewed as a methodological choice. Cosmologists may choose to incorporate maximum entropy conditions when formulating models to ensure that they are based on the most probable configurations that the universe could adopt given certain constraints (e.g., energy density, expansion rate). This approach can help in making predictions and understanding the statistical properties of cosmological structures.
\end{itemize}
Thus, any isolated system evolves toward a state of maximum entropy along a convex curve. In other words, the entropy should be a convex function of redshift with a negative sign for the second derivative of entropy.
\begin{align}
	\label{eqn:SH3}
	\underbrace{S''(z) < 0}_{\rm Convexity~ condition}
\end{align}

 \begin{figure}[ht]
	\centering
	\includegraphics[width=16 cm]{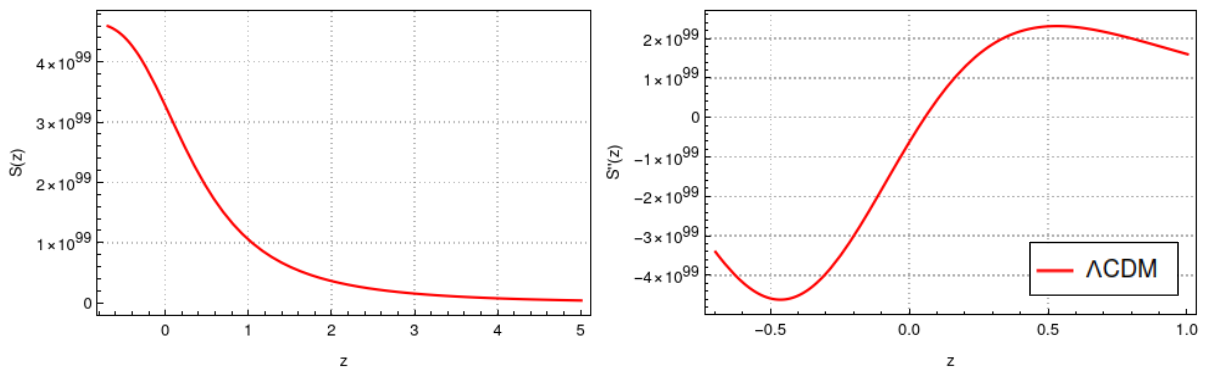}
	\caption{\footnotesize The entropy and maximum entropy test of $\Lambda$CDM model.} 
	\label{DLCDM}
\end{figure}
\subsection{Convexity Problem in the Standard Model through Additive and Non-additive Entropy}
Fig. \ref{DLCDM} illustrates that while the $\Lambda$CDM standard model aligns with the principle of increasing entropy over time, the second derivative of entropy exhibits both a maximum and a minimum at late times, conflicting with the maximum entropy condition. In the following, we will explore two approaches to address this issue: first, the Tsallis generalization of entropy, and second, the effective quadratic model.

\begin{figure}[ht]
	\centering
	\includegraphics[width=10 cm]{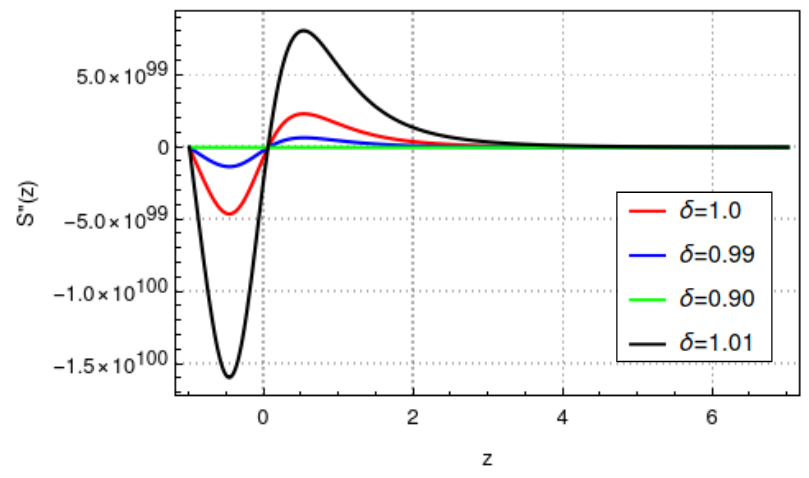}
	\caption{\footnotesize The maximum entropy condition for $\Lambda$CDM ($\delta=1.0$) and Tsallis form of it ($\delta\ne 1.0$).} 
	\label{TLCDM}
\end{figure}

The generalized Tsallis non-additive form of the entropy is given by $S_H=\gamma A_{\rm H}^{\delta}$, where $\delta$ is the non-extensive parameter.  It is obvious that the
area law of entropy is restored for $\delta=1$ and $\gamma = \frac{k_B}{4l_p^2}$. Fig. \ref{TLCDM} shows that applying Tsallis entropy does not address the maximum entropy problem in the standard model. It merely alters the range of maximum and minimum values, shifting their amplitudes higher or lower compared to additive entropy, depending on the $\delta$ parameter changes.
\subsection{Convexity of Quadratic Model and Consistency with DESI DR2}
\begin{figure}[ht]
	\centering
	\includegraphics[width=10 cm]{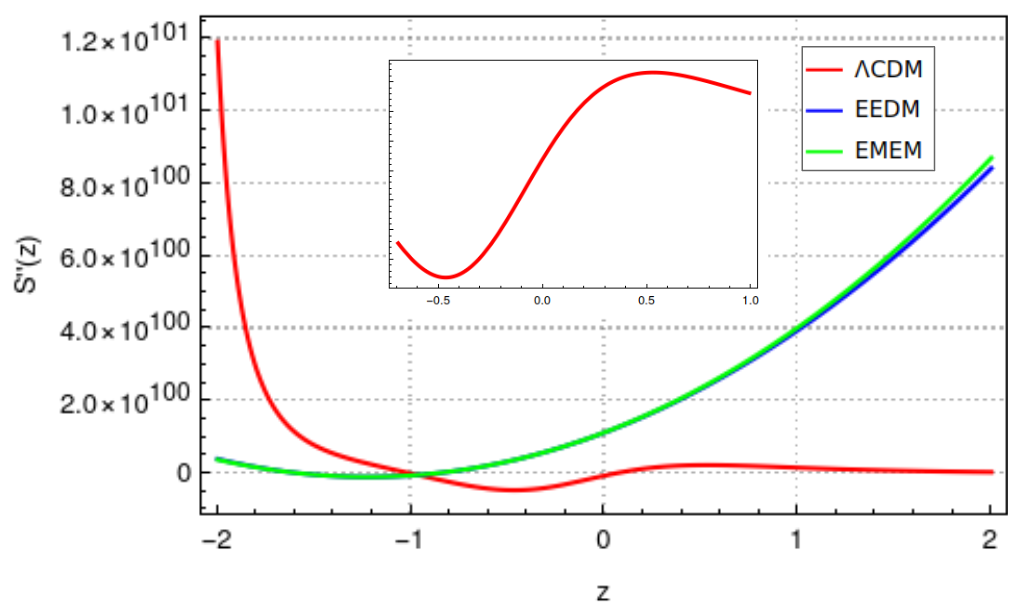}
	\caption{\footnotesize The negative value of $S''(z)$ (maximum entropy) is presented in the future for three models.}
	\label{DLCDM2}
\end{figure}

The right plots from the bottom row in Figures \ref{DDSV} and \ref{DDSCV} indicate that the convexity problem is absent in models that incorporate the evolving of dark energy, which adhere to the maximum entropy condition. Fig. \ref{DLCDM2} presents a comparative diagram of the second derivative of entropy across all models. It shown that in models dominated by effective dark matter (Fig. \ref{DDSC}) exhibit a decreasing trend in entropy until reaching their respective minima. However, this issue occurs in the opposite direction in models dominated by effective dark energy Fig. \ref{DDSV}, leading to a cosmic state of maximum entropy.

\textsf{Now the question arises: why should we investigate the condition of maximum entropy in the universe?} From a thermodynamic perspective, if the universe approaches a state of maximum entropy, it implies a transition where entropy ceases to increase and begins to decrease, indicating a significant change in the nature of its evolution. Moreover, in the context of the cosmological models discussed, such a maximum entropy condition would suggest that the ongoing entropy growth could reverse, with the entropy decreasing in the universe's future trajectory. This behavior is intimately connected to the sign of the \(\xi_{\rm de}\) free parameter; if the sign of \(\xi_{\rm de}\) is reversed, the evolution of entropy—initially increasing—may instead decline (see Fig.\ref{SIGN}), signifying a fundamental shift in the thermodynamic arrow of time. 

Interestingly, by considering this sign change of \(\xi_{\rm de}\) as presented in Equation (\ref{slozz}), for the late universe, the model permits not only the non-phantom phase but also the phantom phase and the crossing of the phantom divide. More precisely in present universe in redshift $z\rightarrow0$, we may adopt the approximations \( w_{\rm de} = -1 \pmb{\pm} \xi_{\rm de} \).  This extension enhances the model's compatibility with recent observational constraints \eqref{wobs2}, such as those from DESI DR2~\cite{DESI:2025zgx}, which suggest possible deviations from a pure cosmological constant scenario. Consequently, the integration of this sign-reversal in the \(\xi_{\rm de}\) parameter broadens the scope of cosmological evolution, allowing the model to accommodate both the observed accelerated expansion and the thermodynamic considerations related to maximum entropy, thus aligning with current observational evidence.

\section{Summary and Conclusions}
\label{VI}

In this work, we have systematically investigated the cosmological implications of a quadratic equation of state (EoS) $P = w\rho + b\rho^2$ within a thermodynamic framework. The key results and implications are as follows:

\begin{itemize}
	\item \textbf{Phantom Crossing and Observational Consistency}: 
	The quadratic correction ($b\rho^2$) enables a natural crossing of the phantom divide ($w = -1$), a feature challenging to achieve in linear dark energy models. This behavior aligns with recent observational constraints from DESI DR2 and Planck, particularly the inferred evolution of the dark energy EoS parameter.
	
	\item \textbf{Thermodynamic Evolution}: 
	The model predicts a late-time approach to maximum entropy ($S''(z) < 0$) for specific parameter ranges of the nonlinear coupling coefficient $\xi \equiv b\rho_0$. The sign of $\xi$ determines the direction of the entropy evolution.
	
	\item \textbf{Parameter Significance}: 
	The dimensionless parameter $\xi$ governs the EoS curvature and late-time dynamics. Its redshift-dependent effects allow transitions between phantom ($w < -1$) and non-phantom ($w > -1$) regimes without singularities, offering a versatile alternative to the $\Lambda$CDM paradigm.
\end{itemize}

\noindent
These results demonstrate that the quadratic EoS provides a mathematically consistent framework to address key observational tensions (e.g., phantom crossing) while maintaining thermodynamic viability. Future work could explore the model's predictions for large-scale structure formation and early-universe inflation, where nonlinear EoS terms may play an equally significant role.
\section*{Acknowledgement}
This work has been supported by the Islamic Azad University, Qom Campus (Q.C.), Qom, Iran.
\section*{Data Availability}
There is no new data associated with this article.
\vspace{1cm}

\section*{Acknowledgments}
	This work has been supported by the Islamic Azad University, Qom Branch, Qom, Iran.

	\bibliography{Ashraf_GSL.bib} 
	
\end{document}